\documentclass[aps,twocolumn,a4paper,floatfix,superscriptaddress,pra,longbibliography]{revtex4-1}
\usepackage[toc,page]{appendix}
\usepackage{braket}
\usepackage{epsfig,graphicx,times}
\usepackage{amstext}
\usepackage{amsmath}
\usepackage{amssymb}
\usepackage{mathrsfs}
\usepackage{dcolumn}
\usepackage{bm}
\usepackage[tight]{subfigure}
\usepackage[colorlinks,linkcolor=blue,anchorcolor=blue,citecolor=blue,urlcolor=blue]{hyperref}
\usepackage{color}
\usepackage{siunitx}
\usepackage{appendix}
\usepackage{placeins}
\usepackage{textcomp}
\usepackage{bbold}
\usepackage{float}
\usepackage{verbatim}
\usepackage{minitoc}
\usepackage[dvipsnames]{xcolor}

\usepackage{soul}
\usepackage[framemethod=tikz]{mdframed}
\usepackage{graphicx}
\usepackage{epstopdf}
\usepackage[normalem]{ulem}

\begin{document}

\title{Enhancing Optomechanical Entanglement and Mechanical Squeezing by the Synergistic Effect of Quadratic Optomechanical Coupling and Coherent Feedback}

\author{Ya-Feng Jiao}
\email{yfjiao@zzuli.edu.cn}
\affiliation{School of Electronics and Information, Henan Key Laboratory of Information Functional Materials and Sensing Technology, Zhengzhou University of Light Industry, Zhengzhou 450001, P.R.China}
\affiliation{Academy for Quantum Science and Technology, Zhengzhou University of Light Industry, Zhengzhou 450001, P.R.China}

\author{Ruo-Chen Wang}
\affiliation{School of Electronics and Information, Henan Key Laboratory of Information Functional Materials and Sensing Technology, Zhengzhou University of Light Industry, Zhengzhou 450001, P.R.China}

\author{Jing-Xue Liu}
\affiliation{School of Physics, Henan Normal University, Xinxiang 453007, China}

\author{Huilai Zhang}
\affiliation{School of Electronics and Information, Henan Key Laboratory of Information Functional Materials and Sensing Technology, Zhengzhou University of Light Industry, Zhengzhou 450001, P.R.China}
\affiliation{Academy for Quantum Science and Technology, Zhengzhou University of Light Industry, Zhengzhou 450001, P.R.China}

\author{Ya-Chuan Liang}
\affiliation{School of Electronics and Information, Henan Key Laboratory of Information Functional Materials and Sensing Technology, Zhengzhou University of Light Industry, Zhengzhou 450001, P.R.China}
\affiliation{Academy for Quantum Science and Technology, Zhengzhou University of Light Industry, Zhengzhou 450001, P.R.China}

\author{Yan Wang}
\affiliation{School of Electronics and Information, Henan Key Laboratory of Information Functional Materials and Sensing Technology, Zhengzhou University of Light Industry, Zhengzhou 450001, P.R.China}
\affiliation{Academy for Quantum Science and Technology, Zhengzhou University of Light Industry, Zhengzhou 450001, P.R.China}

\author{Le-Man Kuang}
\affiliation{Key Laboratory of Low-Dimensional Quantum Structures and Quantum Control of Ministry of Education, \\
Department of Physics and Synergetic Innovation Center for Quantum Effects and Applications, \\ Hunan Normal University, Changsha 410081, China}
\affiliation{Academy for Quantum Science and Technology, Zhengzhou University of Light Industry, Zhengzhou 450001, P.R.China}

\author{Hui Jing}
\affiliation{Key Laboratory of Low-Dimensional Quantum Structures and Quantum Control of Ministry of Education, \\
Department of Physics and Synergetic Innovation Center for Quantum Effects and Applications, \\ Hunan Normal University, Changsha 410081, China}
\affiliation{Academy for Quantum Science and Technology, Zhengzhou University of Light Industry, Zhengzhou 450001, P.R.China}
\affiliation{College of Science, NUDT, Changsha 410073, P.R.China}

\date{\today}

\begin{abstract}
In this paper, we investigate how to achieve strong optomechanical entanglement and mechanical squeezing in a membrane-embedded cavity optomechanical system incorporating a coherent feedback loop, where the membrane interacts with the cavity mode through both linear and quadratic optomechanical couplings. This hybrid optomechanical architecture offers a flexible tunability of intrinsic system parameters, thereby enabling controlled stiffening or softening of the mechanical mode through adjusting quadratic optomechanical coupling, as well as effective modulation of the cavity decay rate via feedback control. More importantly, the synergistic interplay effect allows for a strategic reconfiguration of the system's stability regime, which in turn permits the presence of significantly enhanced effective optomechanical coupling strengths before entering the unstable regime. Exploiting these unique features, we showcase that optomechanical entanglement can be substantially enhanced with positive coupling sign and suitable feedback parameters, while strong mechanical squeezing beyond the $3$dB limit is simultaneously achieved over a broad parameter range with negative coupling sign, reaching squeezing degree above $10$dB under optimized conditions. Our proposal, establishing an all-optical method for generating highly entangled or squeezed states in cavity optomechanical systems, opens up a new route to explore macroscopic quantum effects and to advance quantum information processing.
\end{abstract}

\maketitle

\section{Introduction}

Quantum entanglement~\cite{Horodecki2009RMP} and squeezing~\cite{Andersen2016PS}, as striking features of quantum mechanics, have attracted intense interests owing to their potential applications in modern quantum science and technologies~\cite{Dowling2003PTRSA}. In particular, the preparation of entangled or squeezed states in massive mechanical systems has long been an ongoing pursuit~\cite{Braunstein2005RMP,Andersen2010LPR}, which is not only because of their great significance for the fundamental tests of quantum theory and the exploration of the classical-quantum boundary~\cite{Modi2012RMP}, but also because such states can provide indispensable resources for advancing quantum technologies beyond the classical limits, e.g., improving sensitivity in ultraprecision measurement~\cite{Xia2023NP}, enhancing security in communication~\cite{He2015PRL}, and boosting computational efficiency in information processing~\cite{Zhang2018NSR}. Nevertheless, the generation and preservation of such macroscopic nonclassical states are severely hindered by the decoherence effect induced by environmental thermal noise. In the past decades, a lot of effort has been devoted both theoretically and experimentally to overcome this difficulty, leading to a variety of schemes for realizing macroscopic entanglement~\cite{Luo2022PRL} and squeezing~\cite{Esteve2008Nature}. For instance, entanglement generation has been explored through exploiting injection of quantum squeezing~\cite{Wang2013PRL,Jiao2024LPR} or synthetic gauge fields~\cite{Lai2022PRL}, dark-mode~\cite{Lai2022PRR} or feedback control~\cite{Riste2013Nature,Li2017PRA,Miki2023PRA}, high-frequency resonance effect~\cite{Shang2024PRApp}, photon counting~\cite{Ho2018PRL}, and nonreciprocal manipulation~\cite{Jiao2020PRL}.

On the other hand, cavity optomechanical (COM) systems~\cite{Aspelmeyer2014RMP}, capable of cooling massive mechanical oscillators to their ground state~\cite{Rocheleau2010Nature,Connell2010Nature,Whittle2021Science}, have emerged as a versatile platform for exploring a wide range of nonclassical effects~\cite{Xiong2015SCPMA}, including single-photon or single-phonon blockade~\cite{Rabl2011PRL}, quantum phase transitions~\cite{Lu2018PRAPP}, quantum chaos~\cite{Zhu2019PRA}, phonon lasing~\cite{Sheng2020PRL}, and optomechanical Bell tests~\cite{Simon2018PRL}, to name a few. In a recent experiment, nonclassical correlations were even produced between light and $40\,\textrm{kg}$ mirrors~\cite{Yu2020Nature}, showing a joint quantum uncertainty below the standard quantum limit. A closely related research topic to the present study is the generation and manipulation of macroscopic entanglement~\cite{Vitali2007PRL,Genes2008PRA,Ghobadi2014PRL} and squeezing~\cite{Wollman2015Science,Agarwal2016PRA,Marti2024NP} involving massive mechanical oscillators. Recently, by exploiting the down-conversion interaction enabled by the radiation-pressure-induced nonlinear COM coupling~\cite{Aspelmeyer2014RMP,Xiong2015SCPMA}, remarkable progress has been made towards the observation of quantum entanglement between light and motion~\cite{Riedinger2016Nature,Palomaki2013Science}, between propagating optical fields~\cite{Chen2020NC,Barzanjeh2019Nature}, and between massive mechanical oscillators~\cite{Kotler2021Science,Mercier2021Science}. Meanwhile, by effectively reducing the quantum noise below the standard quantum limit, COM system also offers a powerful platform for generating strong optical and mechanical squeezing beyond the $3$-dB limit~\cite{Dassonneville2021PRXQ,Lei2016PRL,Kamba2025Science}, which is highly beneficial for high-precision quantum sensing applications.

In this work, we investigate how to achieve coherent enhancement of COM entanglement and mechanical squeezing through using the synergistic effect of quadratic optomechanical coupling and coherent feedback. On one hand, the feasibility of COM systems supporting both linear and quadratic couplings (LOC and QOC) has been demonstrated across a variety of experimental platforms, including dielectric-membrane-embedded resonators~\cite{Thompson2008Nature,Sankey2010NP,Vanner2011PRX,Flowers2012APL}, trapped cold atoms~\cite{Purdy2010PRL} or microspheres~\cite{Li2011NP}, microwave superconducting circuits~\cite{Rocheleau2010Nature}, double-disk structures~\cite{Lin2009PRL}, and magnomechanical devices~\cite{Makinen2025NC}. The presence of QOC introduces a static mechanical response sensitive to the interaction sign~\cite{Zhang2014SCPMA}, thus offering a versatile mechanism for mechanical frequency modulation. Leveraging this unique degree of freedom, such hybrid systems can exhibit superior performance over pure LOC configurations, leading to enhanced mechanical squeezing and cooling~\cite{Rocheleau2010Nature,Xuereb2013PRA}, efficient optical harmonic generation~\cite{Zhang2014PRA}, and more robust optomechanical entanglement~\cite{Ghorbani2025PRA}. Notably, the inherent tunability of such COM systems featuring dual couplings has recently facilitated the realization of a continuous time crystal coupled to a mechanical mode, showing that the time-crystal frequency can be modulated by mechanical motion to access diverse optomechanical regimes~\cite{Makinen2025NC}. On the other hand, coherent feedback control has recently emerged as a promising technique because it bypasses noisy measurements, thereby preserving the quantum coherence of the signals mediating the feedback. This approach provides a robust framework for quantum state engineering, as evidenced by recent works demonstrating ground-state cooling across a broad range of parameters~\cite{Du2025PRA}, the enhancement of few-photon optomechanical effects~\cite{Wei2021OE}, the generation of strong optical or mechanical squeezing~\cite{Xiong2020PR,Bemani2024PRApp}, and the effective preservation of quantum coherence~\cite{Mekonnen2024PS}.
\begin{figure}[htbp]
\centering
\includegraphics[width=0.48\textwidth]{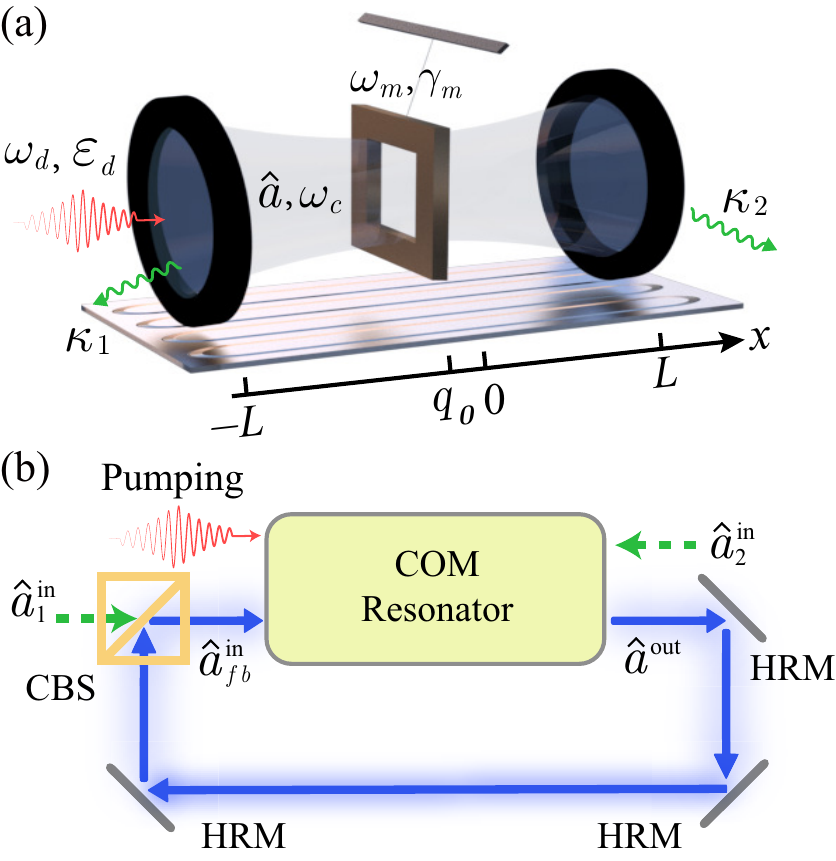}
\caption{Schematic diagram of a membrane-embedded COM system coupled with a coherent feedback loop. (a) The COM system comprises a FP cavity formed by two fixed mirrors with optical decay rates $\kappa_{1}$ and $\kappa_{2}$, inside which a partially reflecting membrane with reflectivity $R_{m}$ is located near the cavity center. The FP cavity with resonance frequency $\omega_{c}$ is driven by a coherent laser field of frequency $\omega_{d}$ and amplitude $\varepsilon_{d}$. The membrane, characterized by resonance frequency $\omega_{m}$ and damping rate $\gamma_{m}$, couples to the cavity mode through both linear ($g_{1}$) and quadratic ($g_{2}$) optomechanical couplings. (b) The coherent feedback loop consists of three highly reflected mirrors (HRMs) and a controllable beam splitter (CBS) with tunable reflection coefficient $r_B$. The blue arrows indicate that the optical output field $\hat{a}^{\rm out}$ transmitted from the right-hand mirror are fed back to the cavity through the left-hand mirror. $\hat a_{1}^{\mathrm{in}}$ and $\hat a_{2}^{\mathrm{in}}$ describe the optical input noise arising from the zero-point fluctuations in the vacuum entering the cavity from the CBS and the right-hand mirror.}\label{Fig1}
\end{figure}

Specifically, we here show that by harnessing the interplay of QOC and coherent feedback, both the effective mechanical frequency and cavity decay rate become highly tunable, thus allowing for a strategic reconfiguration of the system's stability regime. By optimizing the controlling parameters, the instability threshold can be shifted, enabling the system to sustain significantly stronger effective optomechanical coupling strengths before entering the unstable regime. This synergistic modulation of system stability is otherwise unattainable by using either QOC or coherent feedback alone. Exploiting this feature, it is found that COM entanglement can be considerably enhanced for positive QOC with suitable feedback parameters, reaching an enhancement factor of about $5$ under optimized conditions. In addition, strong mechanical squeezing beyond the $3$dB limit is simultaneously achieved for negative QOC, with optimal squeezing degrees exceeding $9$dB under proper feedback parameters. Overall, our proposed scheme, showcasing the transformative potential for engineering and improving various nonclassical effects involving massive mechanical systems~\cite{Rabl2011PRL,Lu2018PRAPP,Zhu2019PRA,Simon2018PRL,Sheng2020PRL}, is expected to advance a wide range of COM-based applications~\cite{Dowling2003PTRSA,Aspelmeyer2014RMP,Xiong2015SCPMA} ranging from quantum sensing~\cite{Xia2023NP} to quantum networking~\cite{He2015PRL} and quantum computing~\cite{Zhang2018NSR}.

This paper is structured as follows. In Sec.\,\ref{secII}, we introduce the theoretical model of the proposed COM system and derive the effective Hamiltonian, on the basis of which the system dynamics and the quantitative measures of COM entanglement and mechanical squeezing are obtained. In Secs.\,\ref{secIII} and \ref{secV}, we present the numerical simulations of the behavior of COM entanglement and mechanical squeezing under various controlling parameters, and analyze the underlying physical mechanism and experimental feasibility. In Sec.\,\ref{secVI}, we provide a brief summary of the main results.

\section{Theoretical model}\label{secII}
As shown in Fig.\,\ref{Fig1}(a), we consider a hybrid COM system consisting of a Fabry-Perot (FP) cavity and a membrane with finite reflectivity $R_{m}$. The FP cavity is formed by two fixed mirrors located at positions $x=\pm L$, and the membrane, with its thickness much smaller than the optical wavelength, is placed inside the cavity at an equilibrium position $x=q_{0}$. In this configuration, the mode frequencies of the FP cavity and the type of optomechanical interactions are determined by the value of $R_{m}$ and $q_{0}$. On one hand, in the case of $R_{m}=1$ and $q_{0}=0$, the FP cavity is effectively divided into two subcavities, supporting two-fold degenerate optical modes at frequency $\omega_{n}=n\pi c/L$, where $c$ is the speed of light, $n=2L/\lambda_{n}$ is the mode number, and $\lambda_{n}=2\pi c/\omega_{n}$ denotes the corresponding optical wavelength. On the other hand, when $R_{m}\neq1$ and $q_{0}\neq0$, the optical degeneracy of the two subcavity modes is lifted, yielding a pair of non-degenerate modes at frequencies $\omega_{n,e}$ and $\omega_{n,o}$, which correspond to the even and odd half-wavelength modes of the full FP cavity~\cite{Bhattacharya2008PRA}, respectively. For large mode numbers with $L\gg\lambda_n$ and $q_0\ll\lambda_n$, the round-trip time of light is approximately the same for both subcavity modes~\cite{Bhattacharya2008PRA}, i.e., $\tau=2L/c$. Since the membrane's mechanical motion is much slower than the intracavity field dynamics, satisfying $\tau \ll 1/\omega_m$, the cavity resonance frequencies $\omega_{n,e}$ and $\omega_{n,o}$ follow the membrane motion adiabatically and can thus be treated as instantaneous, position-dependent functions of the membrane displacement $q_1$. Particularly, when the membrane is placed near the middle of the cavity (i.e., $q_{0}\approx0$, near an antinode), the cavity resonance frequency is primarily dominated by $\omega_{n,o}$ of the odd half-wavelength modes, which is given by~\cite{Bhattacharya2008PRA}
\begin{align}
\omega_{n,o}(q_{1})\simeq&\omega_{n}+\dfrac{\pi}{\tau}\!-\!\dfrac{1}{\tau}\left.\{\sin^{-1}[\sqrt{R_{m}}\cos(2k_{n}q_{1})]\right.
\notag\\
&\left.+\sin^{-1}(\sqrt{R_{m}})\right.\},
\end{align}
where $k_{n}=\omega_{n}/c$. Given that the equilibrium position $q_{0}$ of the membrane is small, we expand $\omega_{n,o}(q_{1})$ in powers of $q_{0}$ up to the second order,
\begin{align}
\omega_{n,o}(q_{1})=&\omega_{n,o}(q_{0})+\left.\dfrac{d\omega_{n,o}(q_{1})}{dq_{1}}\right|_{q_{1}=q_{0}}(q_{1}-q_{0})
\notag\\
&+\dfrac{1}{2}\left.\dfrac{d^{2}\omega_{n,o}(q_{1})}{dq_{1}^{2}}\right|_{q_{1}=q_{0}}(q_{1}-q_{0})^{2}
\notag\\
&=\omega_{c}+g_{1}q+g_{2}q^{2}, \label{eq:w_no}
\end{align}
with
\begin{align}
\omega_{c}=&\omega_{n,o}(q_{0}),
\notag\\
g_{1}=&\dfrac{2k_{n}}{\tau\sigma}\sqrt{R_{m}}\sin(2k_{n}q_{0}),
\notag\\
g_{2}=&\dfrac{2k_{n}^{2}}{\tau\sigma^{3}}\sqrt{R_{m}}(1-R_{m})\cos(2k_{n}q_{0}),
\notag\\
\sigma=&\sqrt{\sin^{2}(2k_{n}q_{0})+(1-R_{m})\cos^{2}(2k_{n}q_{0})}, \label{eq:g12}
\end{align}
where $q=q_{1}-q_{0}$ is the displacement of the membrane from its equilibrium position. Equations (\ref{eq:w_no}) and (\ref{eq:g12}) indicate that the sign and magnitude of the LOC and QOC strengths $g_{1}$ and $g_{2}$ are explicitly determined by the equilibrium position $q_{0}$ of the membrane inside the cavity. In particular, when the membrane is placed in the middle of the cavity (i.e., $q_{0}=0$), the optomechanical coupling is purely quadratic, with $g_{1}=0$ and $g_{2}\neq0$. In contrast, when the membrane has a slight mechanical displacement from the middle (i.e., $q_{0}\approx0$), both LOC and QOC are present, with $g_{1}\neq0$ and $g_{2}\neq0$. In this paper, we focus on the latter case for two reasons: (i) our work aims to investigate the role of QOC in enhancing optomechanical entanglement and mechanical squeezing, and (ii) the latter case is more general since placing the membrane exactly at the middle of the cavity is experimentally challenging in practice. Accordingly, the Hamiltonian of this COM system reads
\begin{align}
H=&\hbar\omega_{c}\hat{a}^{\dagger}\hat{a}+\dfrac{\hbar\omega_{m}}{2}(\hat{p}^{2}+\hat{q}^{2})+\hbar g_{1}\hat{a}^{\dagger}\hat{a}\hat{q}+\hbar g_{2}\hat{a}^{\dagger}\hat{a}\hat{q}^{2}\notag\\
&+i\hbar{\varepsilon}_{d}(e^{-i\omega_{d}{t}}\hat{a}^{\dagger}-e^{i\omega_{d}{t}}\hat{a}), \label{eq:Hamiltonian}
\end{align}
where $\hat{a}$ ($\hat{a}^{\dagger}$) denotes the annihilation (creation) operator of the cavity mode, and $\hat{q}$ and $\hat{p}$ are the dimensionless position and momentum operators of the membrane, respectively. In Eq.\,(\ref{eq:Hamiltonian}), the first two terms correspond to the free Hamiltonians of the cavity and the mechanical modes, respectively. The third and fourth terms describe the LOC and QOC between the cavity and the membrane, with coupling strengths $g_{1}$ and $g_{2}$ [see Eq.\,(\ref{eq:g12})]. In our model, the role of LOC is to enable the down-conversion interaction between the optical and mechanical modes and thereby generate COM entanglement~\cite{Vitali2007PRL}, while the QOC is used to provide an additional degree of freedom to modulate COM entanglement through tuning the effective mechanical frequency. The last term is the Hamiltonian of the coherent driving field with amplitude $\varepsilon_{d}$ and frequency $\omega_{d}$. $\varepsilon_{d}$ is related to the input laser power $P_{d}$ by $|\varepsilon_{d}|=\sqrt{2P_{d}\kappa_{1}/\hbar \omega_{d}}$, with $\kappa_{1}$ the cavity decay rate due to optical transmission through the left-hand mirror.

By considering the system dissipations and environmental input noises, the dynamical evolution of this COM system can be fully characterized by the quantum Langevin equations (QLEs):
\begin{align}
&\dot{\hat{q}}=\omega_{m}\hat{p}, \notag \\
&\dot{\hat{p}}=-\omega_{m}\hat{q}-\gamma_{m}\hat{p}-g_{1}\hat{a}^{\dagger}\hat{a}-2g_{2}\hat{a}^{\dagger}\hat{a}\hat{q}+\hat{\xi}, \notag \\
&\dot{\hat{a}}=-(i\Delta_{c}+\kappa_{1}+\kappa_{2})\hat{a}-ig_{1}\hat{a}\hat{q}-ig_{2}\hat{a}\hat{q}^{2}+\varepsilon_{d}\notag\\
&~~~~~~~+\sqrt{2\kappa_{1}}\hat{a}_{1}^{\textrm{in}}
+\sqrt{2\kappa_{2}}\hat{a}_{2}^{\textrm{in}}, \label{eq:QLEs}
\end{align}
where $\Delta_{c}=\omega_{c}-\omega_{d}$, $\gamma_m$ denotes the mechanical damping rate, and $\kappa_{2}$ denotes the cavity
decay rate of the right-hand mirror. $\hat a_{1}^{\mathrm{in}}$ and $\hat a_{2}^{\mathrm{in}}$ describe the optical input noise arising from the zero-point fluctuations in the vacuum entering the cavity from the CBS and the right-hand mirror, which have zero mean and are characterized by the following nonvanishing correlation function~\cite{Zoller2000book}: $\langle \hat{a}_{j}^{\textrm{in}}(t)\hat{a}_{j}^{\textrm{in},\dagger}(t')\rangle=\delta(t-t')$, with $j=1,2$. $\hat{\xi}$ is the Brownian thermal noise operator, which describes the stochastic Brownian force acting on the membrane. The correlation function of $\hat{\xi}$ is typically not delta-correlated, i.e.,
$\langle\hat{\xi}(t) \hat{\xi}(t^{\prime})\rangle=\frac{\gamma_m}{\omega_m}\int\frac{d\omega}{2\pi}\mathrm{e}^{-\mathrm{i} \omega(t-t^{\prime})}{\omega}[\operatorname{coth}(\frac{\hbar \omega}{2 k_{\mathrm{B}}T})+1]$,
which describes a non-Markovian process. However, for high-Q mechanical membranes, i.e., $Q_{m}=\omega_{m}/\gamma_{m}\gg1$, one can safely make the Markovian approximation and the correlation function of $\hat{\xi}$ can be reduced to a delta-correlated form~\cite{Zoller2000book}: $\langle\hat{\xi}(t)\hat{\xi}(t')\rangle\simeq\gamma_{m}(2\bar{n}_{m}+1)\delta(t-t')$,
where $\bar{n}_{m}\!=\![\exp(\hbar\omega_{m}/k_{\textit{B}}\mathit{T})-1]^{-1}$ is the equilibrium mean thermal phonon number, with $k_{\textit{B}}$ the Boltzmann constant and $\mathit{T}$ the bath temperature of the mechanical mode.
\begin{figure*}[htbp]
\centering
\includegraphics[width=0.96\textwidth]{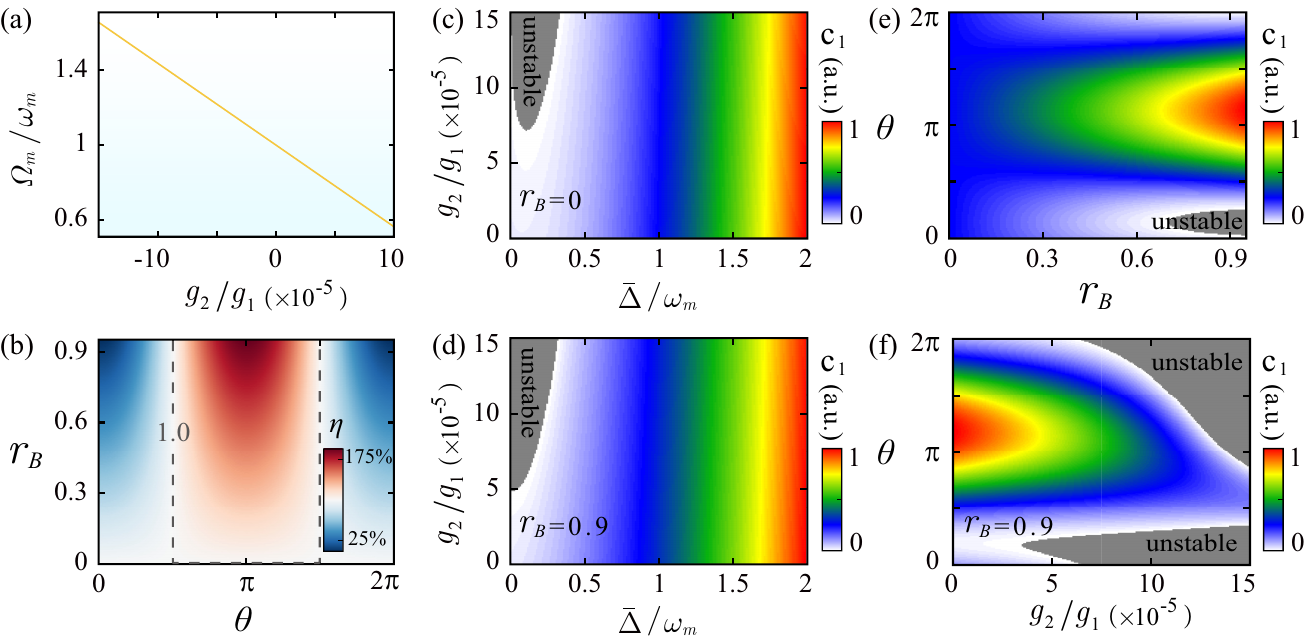}
\caption{Synergistic effects of QOC and coherent feedback on the system parameters and stability condition. (a) The effective membrane frequency $\Omega_{m}$ versus the QOC strength $g_{2}/g_{1}$, with $\bar{\Delta}/\omega_{m}=0.25$. When changing the sign of QOC, $\Omega_{m}$ increases for $g_{2}/g_{1}<0$ and decreases for $g_{2}/g_{1}>0$, which provides an efficient method for the modulation of mechanical frequency. (b) The decay ratio $\eta$, defined by $\eta\equiv\tilde\kappa/(\kappa_{1}+\kappa_{2})$, is plotted as a function of the feedback parameters $r_{B}$ and $\theta$. When adjusting $r_{B}$ and $\theta$, $\eta$ is amplified or reduced in certain parameter regions, indicating that the cavity decay rate can be effectively tuned by coherent feedback. (c,d) Density plot of the stability condition $C_{1}$ as a function of the scaled optical detuning $\bar{\Delta}/\omega_{m}$ and the QOC strength $g_{2}/g_{1}$, corresponding to (c) $r_{B}=0$ and (d) $r_{B}=0.9$, $\theta=3\pi/2$. (e) Density plot of the stability condition $C_{1}$ versus the reflection coefficient $r_{B}$ and the optical phase shift $\theta$, with $\bar{\Delta}/\omega_{m}=0.25$ and $g_{2}/g_{1}=6\times10^{-5}$. (f) Density plot of the stability condition $C_{1}$ versus the QOC strength $g_{2}/g_{1}$ and the optical phase shift $\theta$, with $\bar{\Delta}/\omega_{m}=0.25$ and $r_{B}=0.9$. The default parameters used here are $\kappa_1=\kappa_2=2\pi\times1.5\,\mathrm{MHz}$, $\omega_m=2\pi\times10\,\mathrm{MHz}$, $g_{1}=-1351.38\,\mathrm{Hz}$, $\lambda_{d}=810\,\mathrm{nm}$, and $P_{d}=0.8\,\mathrm{mW}$.}\label{Fig2}
\end{figure*}

The dynamics of QLEs\,(\ref{eq:QLEs}) involve radiation-pressure-induced nonlinear COM interactions between the cavity field and the membrane, and thus are difficult to be directly solved. In order to find solutions to QLEs\,(\ref{eq:QLEs}), one can linearize the system dynamics under a sufficiently strong driving condition by expanding each operator as a sum of its steady-state mean value and a quantum fluctuation around it: $\hat{a}=\alpha_{s}+\delta\hat{a}$, $\hat{q}=q_{s}+\delta\hat{q}$, $\hat{p}=p_{s}+\delta\hat{p}$. Inserting this assumption into the QLEs\,(\ref{eq:QLEs}) yields a set of nonlinear algebraic equations for the steady-state mean values and a set of linearized QLEs for the fluctuation operators. By solving the nonlinear algebraic equations, the steady-state mean values are obtained as
\begin{align}
p_{s} =& 0, \notag \\
q_{s} =& \dfrac{-g_{1}|\alpha_{s}|^{2}}{\omega_{m}+2g_{2}|\alpha_s|^{2}},\notag \\
\alpha_{s} =& \dfrac{\varepsilon_{d}}{i\bar{\Delta}+\kappa_{1}+\kappa_{2}},
\end{align}
where
\begin{align}
\bar{\Delta}=\Delta_{c}+g_{1}q_{s}+g_{2}q_{s}^{2} \label{eq:Delta_eff}
\end{align}
is the effective optical detuning including the optomechanically induced frequency shifts. Furthermore, by neglecting the high-order terms in the QLEs for fluctuation operators, i.e., $\delta\hat{a}^{\dagger}\delta\hat{a}$, $\delta\hat{a}^{\dagger}\delta\hat{q}$, $\delta\hat{a}\delta\hat{q}$, $\delta\hat{q}\delta\hat{q}$, and $\delta\hat{a}^{\dagger}\delta\hat{a}\delta\hat{q}$, the linearized QLEs are derived as
\begin{align}
\delta\dot{\hat{q}}=&\ \omega_{m}\delta\hat{p}, \notag \\
\delta\dot{\hat{p}}=&-\Omega_{m}\delta\hat{q}-\gamma_{m}\delta\hat{p}
-\tilde{G}(\delta\hat{a}+\delta\hat{a}^{\dagger})+\hat{\xi}, \notag \\
\delta\dot{\hat{a}}=&-(i\bar{\Delta}+\kappa_{1}+\kappa_{2})\delta\hat{a}-i\tilde{G}\delta\hat{q}
+\sqrt{2\kappa_{1}}\hat{a}_{1}^{\textrm{in}}+\sqrt{2\kappa_{2}}\hat{a}_{2}^{\textrm{in}},
\label{eq:linQLEs}
\end{align}
where $\Omega_{m}=\omega_{m}+2g_{2}|\alpha_s|^{2}$ is the effective membrane frequency, and $\tilde{G}=(g_{1}+2g_{2}q_{s})\alpha_{s}$ denotes the effective COM coupling strength. Here $\alpha_{s}$ is assumed to be real by choosing a suitable phase reference for the cavity field. Note that the validity of this linearization procedure relies on the strong driving condition and the weak single-photon coupling regime $(g_1, g_2)/(\kappa_1+\kappa_2) \ll 1$, which enables that the steady-state mean values are much larger than the fluctuation terms, i.e., $|\alpha_{s}|\gg |\delta \hat{a}|$ and $|q_{s}|\gg|\delta \hat{q}|$. For our considered parameters, we have confirmed that the steady-state photon number $|\alpha_s|^2$ is approximately eight orders of magnitude larger than the mean values of the fluctuation correlations (e.g., $|\langle \delta \hat{a}^{\dagger} \delta \hat{a} \rangle|$ and $|\langle \delta \hat{a}^{\dagger} \delta \hat{q} \rangle|$). This result ensures that the contribution of the neglected higher-order terms in Eq.\,(\ref{eq:linQLEs}) is insufficient to influence the system's dynamics, which is consistent with the standard treatment in weak-coupling optomechanics~\cite{Nunnenkamp2011PRL}. As shown in Fig.\,\ref{Fig2}(a), the effective membrane frequency $\Omega_{m}$ is sensitive to the sign of QOC strength $g_{2}$, namely, $\Omega_{m}$ is enhanced for $g_{2}<0$ and is suppressed for $g_{2}>0$, implying that the membrane becomes either stiffer or softer depending on the sign of $g_{2}$. This tunability of mechanical frequency provides a useful tool for the manipulation of the stability of COM system as discussed below.

When coupling a coherent feedback loop to the COM setup, as depicted in Fig.\,\ref{Fig1}(b), the optical output field transmitted from the right-hand mirror can be sent back into the cavity through the left-hand mirror. By using the standard input-output relation, the output field is given by~\cite{Agarwal2013book}
\begin{align}
\hat{a}^{\rm out}=\sqrt{2\kappa_2} \delta\hat{a} - \hat{a}_{2}^{\rm in}.
\label{Cavityout}
\end{align}
When the optical path between the two mirrors (loop length) is relatively short, the associated time delay of the output field becomes negligible compared with the cavity lifetime, so that the feedback process can be regarded as instantaneous. This is a good approximation for high-Q FP cavities. For example, as shown in Ref.\,\cite{Li2017PRA}, for a $5$-cm FP cavity with a $10$-cm feedback loop, the delay time is about $10^{-10}\,\textrm{s}$, which is much shorter than the typical cavity lifetime ($\sim10^{-7}\,\textrm{s}$, corresponding to an optical Q-factor of $\sim10^{6}$). Particularly, with the optical Q-factor of $Q\sim10^8$ adopted in our work, the cavity lifetime is $\tau = Q/\omega_c \approx 4 \times 10^{-8}$ s. This is still two orders of magnitude longer than the delay time ($\sim10^{-10}$ s) associated with a centimeter-scale feedback loop, thereby ensuring that the instantaneous feedback approximation remains highly accurate for our considered parameters. Accordingly, the new input field modified by the feedback can be modeled as the superposition of the original input noise and the returned output field, which are mixed in a lossless CBS before entering the cavity; the corresponding input field operator is given by
\begin{equation}
\hat{a}_{fb}^{\rm in}=r_{B}e^{i \theta}\,\hat{a}^{\rm out} +t_{B}\hat{a}_1^{\rm in},
\label{BSout}
\end{equation}
where $r_B$ and $t_B$ denote the reflection and transmission coefficients of the CBS, with $r_B^2+t_B^2=1$. The additional optical phase shift $\theta$ of the output field accounts for the accumulated phase delay from light propagation in the feedback loop and is defined as $\theta =2\pi nl/\lambda_{d}$, with $n$ ($l$) the refractive index (length) of the loop and $\lambda_{d}$ the light wavelength. Note that, since both end mirrors of FP cavity are fixed and the membrane oscillates only inside the cavity, the optical path length of the external feedback loop could not be modified by the membrane motion. Consequently, the propagation-induced phase $\theta$ acquired in the feedback loop is independent of the membrane oscillation. Moreover, we also emphasize that the reflection coefficient $r_B$ and additional optical phase shift $\theta$ are defined in an effective way, such that they already incorporate all optical losses and propagation-induced phase shifts in the feedback loop. For example, although both highly reflected mirrors (HRMs) and controllable beam splitters (CBS) may introduce optical dissipation to feedback loop, these loss channels are effectively mapped onto a single equivalent loss channel associated with the CBS denoted by $\hat{a}_1^{\rm in}$. As a result, $r_B$ characterizes the feedback efficiency of the output field, which is limited by the total loop losses, such that $0\leq r_B<1$.

Then, by replacing the bare input noise operator $\hat{a}_1^{\rm in}$ in Eq.\,(\ref{eq:linQLEs}) with the feedback-modified operator $\hat{a}_{fb}^{\rm in}$, we obtain the QLE for the cavity mode in the presence of a coherent feedback loop as
\begin{align}
\delta \dot{\hat{a}}=-(i\tilde\Delta+\tilde\kappa)\delta\hat{a}-i\tilde{G}\delta\hat{q}+\sqrt{2\,\tilde\kappa}\hat A^{\rm in},
\end{align}
where $\tilde\Delta=\bar{\Delta}-2\sqrt{\kappa_{1}\kappa_{2}}r_{B}\sin\theta$ and $\tilde\kappa=\kappa_{1}+\kappa_{2}-2\sqrt{\kappa_{1}\kappa_{2}}r_{B}\cos\theta$ are the feedback modified effective cavity detuning and decay rate, respectively. Moreover, the feedback modified input noise operator $\hat A^{\rm in}$ is given by
\begin{align}
\hat A^{\rm in}=\dfrac{1}{\sqrt{\tilde\kappa}}\left[\left(\sqrt{\kappa_{2}}-\sqrt{\kappa_{1}}r_{B}e^{i\theta}\right)\hat{a}_2^{\rm in}+\!\!\sqrt{\kappa_1}\ t_{B} \,\hat{a}_1^{\rm in}\right],
\end{align}
which corresponds to vacuum noise and obeys the correlation function: $\langle\hat A^{\rm in}(t)\hat A^{\rm in,\dagger}(t')\rangle=\delta(t-t')$. Obviously, in the presence of coherent feedback, the cavity parameters $\tilde\Delta$ and $\tilde\kappa$ become tunable, which can either be enhanced or suppressed by adjusting the feedback parameters $r_{B}$ and $\theta$. To quantify the impact of the coherent feedback on the effective cavity decay rate, we define a decay ratio
\begin{align}
\eta\equiv\dfrac{\tilde\kappa}{\kappa_{1}+\kappa_{2}},
\end{align}
where $\eta<1$ ($\eta>1$) indicates a reduction (enhancement) of the cavity decay rate induced by coherent feedback. In Figure\,\ref{Fig2}(b) shows the dependence of $\eta$ on $r_{B}$ and $\theta$ under $\kappa_{1}=\kappa_{2}$. It is seen that by properly tuning $r_{B}$ and $\theta$, the effective cavity decay rate $\tilde\kappa$ can be significantly reduced, even reaching decay ratio below $25\%$ of the original decay rate. This ability to reduce cavity decay rate is beneficial for enhancing and preserving COM entanglement and mechanical quadrature squeezing as discussed in the following section.

Accordingly, by introducing the optical quadrature operators
$\delta\hat{X}\equiv(\delta\hat{a}^{\dagger}+\delta\hat{a})/\sqrt{2}$ and $\delta
\hat{Y}\equiv i(\delta\hat{a}^{\dagger}-\delta\hat{a})/\sqrt{2}$, together with the corresponding Hermitian input noise operators
$\hat{X}^{\rm in}\equiv(\hat{A}^{\rm in,\dagger}+\hat{A}^{\rm in})/\sqrt{2}$ and
$\hat{Y}^{\rm in}\equiv i(\hat{A}^{\rm in,\dagger}-\hat{A}^{\rm in})/\sqrt{2}$, the feedback-modified linearized QLEs can be written in a compact form
\begin{align}
\dot{u}(t) = A\hat{u}(t)+\hat{n}(t), \label{eq:linQLEs_QO}
\end{align}
where
\begin{align}
\hat{u}(t)&=(\delta\hat{q}, \delta\hat{p}, \delta\hat{X}, \delta\hat{Y})^{T}, \notag\\
\hat{n}(t)&=(0,\hat{\xi},\sqrt{2\tilde\kappa}\hat{X}^{\textrm{in}},\sqrt{2\tilde\kappa}\hat{Y}^{\textrm{in}})^{T},
\end{align}
denote the vectors of quadrature fluctuations and input noises, respectively. The coefficient matrix $A$ is given by
\begin{align}
A=\left(
\begin{matrix}
0 & \omega_{m} & 0 & 0\\
-\Omega_{m} & -\gamma_{m} & \sqrt{2}\tilde{G} & 0\\
0 & 0 & -\tilde\kappa & \tilde\Delta\\
\sqrt{2}\tilde{G} & 0 & -\tilde\Delta & -\tilde\kappa
\end{matrix}
\right).\label{eq:coeffmat}
\end{align}
The system dynamics is stable if and only if all eigenvalues of the coefficient matrix $A$ have negative real parts, which leads to the following two nontrivial stability conditions on the system parameters (see Supporting Information for a detailed derivation):
\begin{align}
&C_{1}=\Omega_m(\tilde\Delta^2+\tilde\kappa^2)-2\tilde{G}^2\tilde\Delta>0,
\notag\\
&C_{2}=2\gamma_m\tilde\kappa
\Big[
\tilde\Delta^4
+\tilde\Delta^2(\gamma_m^2
+2\gamma_m\tilde\kappa
+2\tilde\kappa^2
-2\Omega_m\omega_m)
\nonumber\\
&+(\gamma_m\tilde\kappa
+\tilde\kappa^2
+\Omega_m\omega_m)^2
\Big]
+2\tilde{G}^2\omega_m\tilde\Delta
(\gamma_m+2\tilde\kappa)^2>0.
\label{eq:stability}
\end{align}
In the following, the stability conditions are considered to be satisfied throughout the analysis of entanglement and squeezing. It is also worth noting that, under the red-detuned regime ($\tilde\Delta>0$), the second condition $C_{2}$ is always fulfilled by the system parameters, leaving $C_{1}$ as the only nontrivial constraint. In contrast, for the blue-detuned case ($\tilde\Delta<0$), the situation is reversed and the system stability is entirely determined by $C_2$. In Figs.\,\ref{Fig2}(c)-\ref{Fig2}(f), for the consideration of the red-detuned regime, we plot the dependence of the stability condition $C_1$ on the QOC strength $g_2$ and the feedback parameters $r_B$ and $\theta$. Figures\,\ref{Fig2}(c) and \ref{Fig2}(d) show that compared with the case without feedback ($r_{B}=0$), the presence of feedback considerably broadens the instability region and lowers the instability threshold, showing that instability region is reached at smaller values of $g_2$. Moreover, in the presence of coherent feedback, the system stability is also dependent on the optical phase shift $\theta$ accumulated in the feedback loop. Figure\,\ref{Fig2}(e) shows that although stronger coherent feedback can lead to significant reduction in effective cavity decay rate $\tilde\kappa$ [cf. the parameter regime for $\eta<1$ in Fig.\,\ref{Fig2}(b)], it simultaneously degrades the system stability. Figure\,\ref{Fig2}(f) further demonstrates that, for a fixed value of $r_B$, the system is more stable when $\theta$ is close to $\pi$, whereas it becomes unstable for $\theta$ in vicinity of $0$ or $2\pi$. As well, with the increase of QOC strength $g_2$, the instability region widens in this case. This tunability of the system stability originates from the simultaneous modulation of the intrinsic system parameters $\Omega_m$ and $\tilde\kappa$ by the synergistic effect of QOC and coherent feedback. The influence of this modified stability behavior on COM entanglement and mechanical quadrature squeezing will be further analyzed in the following section.

In the stable regime, owing to the linearized system dynamics and the Gaussian nature of the input noises, the steady state of the COM system, independently of any initial conditions, finally evolves into a zero-mean bipartite Gaussian state, which can be completely characterized by a $4\times4$ covariance matrix (CM) $V$ with its entries defined as
\begin{align}
V_{kl} = & \left\langle\hat{u}_{k}(\infty)\hat{u}_{l}(\infty)\!+\!\hat{u}_{l}(\infty)\hat{u}_{k}(\infty)\right\rangle/2, \notag\\
& k,l=1,2,3,4.
\end{align}
The steady-state CM $V$ can be determined by solving the Lyapunov equation
\begin{align}
	AV+VA^{\text{T}}=-D, \label{eq:Lyapunov}
\end{align}
where $D\!=\!\textrm{Diag}\,[0,\gamma_m (2\bar{n}_{m}+1),\tilde\kappa,\tilde\kappa]$ is the diffusion matrix, and it is defined through $D_{kl}\delta(s-s^{\prime})=\langle\hat{n}_{k}(s)\hat{n}_{l}(s^{\prime})+\hat{n}_{l}(s^{\prime})\hat{n}_{k}(s)\rangle/2$. The Lyapunov equation (\ref{eq:Lyapunov}) is linear for $V$ and can be solved straightforwardly; however, its general exact solution is cumbersome and will not be reported here.
\begin{figure*}[htbp]
\centering
\includegraphics[width=0.96\textwidth]{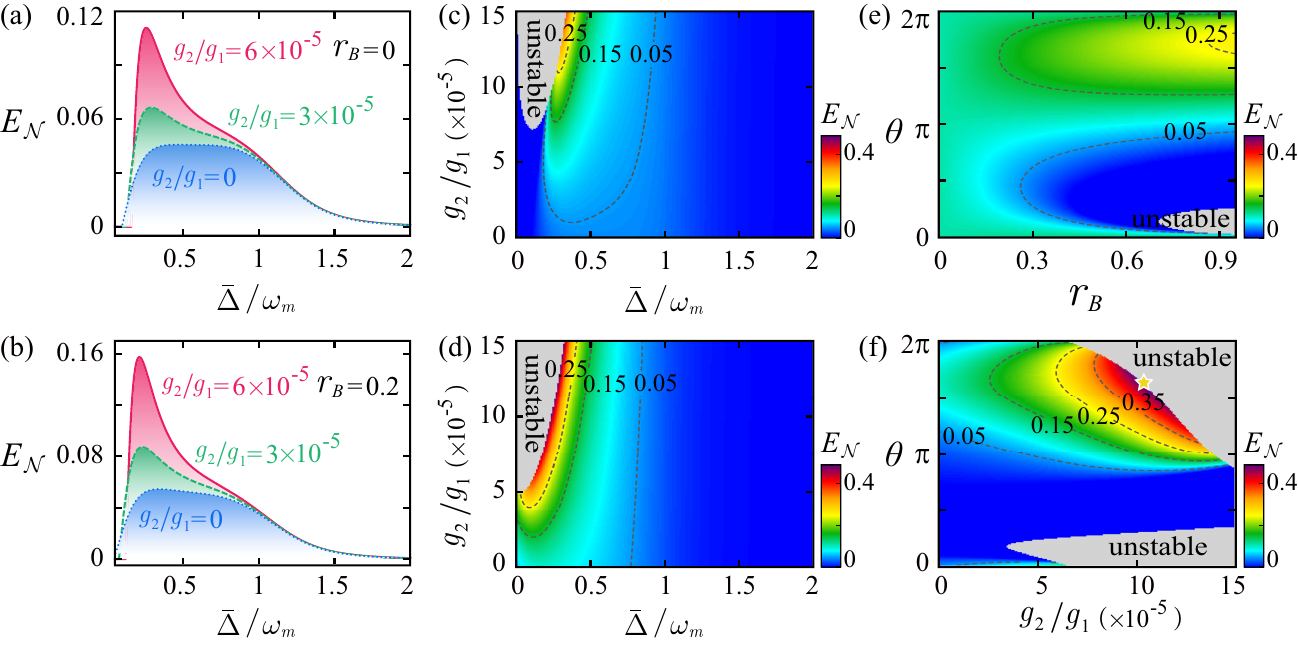}
\caption{Enhancement of COM entanglement via the synergistic effect of QOC and coherent feedback. (a,b) The logarithmic negativity $E_{\mathcal{N}}$ as a function of the scaled optical detuning $\bar{\Delta}/\omega_{m}$ for different values of QOC strength $g_{2}/g_{1}$, showing the case (a) without coherent feedback ($r_{B}=0$) and (b) with coherent feedback ($r_{B}=0.2$, $\theta=3\pi/2$). (c,d) Density plot of the logarithmic negativity $E_{\mathcal{N}}$ verus the scaled optical detuning $\bar{\Delta}/\omega_{m}$ and the QOC strength $g_{2}/g_{1}$, corresponding to (c) $r_{B}=0$ and (d) $r_{B}=0.9$, $\theta=3\pi/2$. (e) Density plot of the logarithmic negativity $E_{\mathcal{N}}$ versus the reflection coefficient $r_{B}$ and the optical phase shift $\theta$, with $\bar{\Delta}/\omega_{m}=0.25$ and $g_{2}/g_{1}=6\times10^{-5}$. (f) Density plot of the logarithmic negativity $E_{\mathcal{N}}$ versus the QOC strength $g_{2}/g_{1}$ and the optical phase shift $\theta$, with $\bar{\Delta}/\omega_{m}=0.25$ and $r_{B}=0.9$. Here we take $\gamma_m=2\pi\times100\,\mathrm{Hz}$, $T=10\,\mathrm{mK}$, and the other parameters are the same as in Fig.\,\ref{Fig2}.}\label{Fig3}
\end{figure*}

\section{The synergistic effect of QOC and coherent feedback on steady-state COM entanglement}\label{secIII}
Entanglement is a key resource for the implementation of various quantum information tasks, thus making its quantification an important problem. In continuous-variable (CV) systems, entanglement can be certified by using different entanglement monotones, among which the logarithmic negativity $E_{\mathcal{N}}$ is widely adopted as a quantitative entanglement measure. For bipartite CV Gaussian systems, the logarithmic negativity $E_{\mathcal{N}}$ is defined as~\cite{Adesso2004PRA}
\begin{align}
E_{\mathcal{N}}=\max\,[0,-\ln (2\nu^{-})], \label{eq:En}
\end{align}
where $\nu^{-}\!=\!2^{-1/2}\{\Sigma(V)-[\Sigma(V)^{2}-4\det\!V]^{1/2}\}^{1/2}$ (with $\Sigma(V)\!\equiv\!\det\mathcal{A}+\det\mathcal{B}-2\det\mathcal{C}$) is the minimum symplectic eigenvalue of the partial transpose of the CM $V$. Here we have rewritten the CM $V$ in a $2\times2$ block form
\begin{align}
V=\left(
\begin{matrix}
\mathcal{A}&\mathcal{C}\\
\mathcal{C}^{\textit{T}}&\mathcal{B}
\end{matrix}
\right),\label{reducedCM}
\end{align}
where the submatrices $\mathcal{A}$ and $\mathcal{B}$ describe the autocorrelations of the optical and mechanical modes, respectively, while the submatrix $\mathcal{C}$ characterizes their cross-correlations. Equation\,(\ref{eq:En}) indicates that the optical and mechanical modes become entangled if and only if $\nu^{-}<1/2$, in which case $E_{\mathcal{N}}$ has a nonzero value. It should be emphasized that $E_{\mathcal{N}}$ quantifies the extent to which the positivity of the partial transpose condition for separability is violated for the Gaussian state, which is equivalent to Simon's necessary and sufficient nonpositive partial transpose criterion (or the related Peres-Horodecki criterion) for bipartite entanglement.

Figure\,\ref{Fig3} shows the synergistic effect of QOC and coherent feedback on COM entanglement, in which the entanglement measure $E_{\mathcal{N}}$ is obtained by numerically solving the Lyapunov equation\,(\ref{eq:Lyapunov}). As shown in Fig.\,\ref{Fig3}(a), to investigate the role of QOC, we first demonstrate the case without coherent feedback ($r_{B}=0$) by plotting the logarithmic negativity $E_{\mathcal{N}}$ as a function of the scaled optical detuning $\bar\Delta/\omega_{m}$ for different values of $g_{2}/g_{1}$. In the absence of QOC, i.e., $g_{2}/g_{1}=0$, $E_{\mathcal{N}}$ is nonvanishing within a finite parameter region around detuning $\bar\Delta/\omega_{m}\simeq0.6$, and the maximum value of $E_{\mathcal{N}}$ is about $0.04$, which means the presence of weak COM entanglement between light and membrane. The slight spectral offset from the nominal COM resonance arises from the radiation-pressure-induced redshift of the cavity frequency [see the last two terms in Eq.\,\ref{eq:Delta_eff}]. In the presence of QOC with the same sign of LOC, $E_{\mathcal{N}}$ exhibits a sharp peak centered at $\bar\Delta/\omega_{m}\simeq0.25$ and it decreases quickly after reaching its maximum value. Besides, larger positive values of $g_{2}/g_{1}$ can lead to enhanced maximum COM entanglement. For example, in the case of $g_{2}/g_{1}=6\times10^{-5}$, the maximum value of $E_{\mathcal{N}}$ is enhanced by approximately $3$ times compared with the case of $g_{2}/g_{1}=0$. In Fig.\,\ref{Fig3}(b), we demonstrate the case with coherent feedback ($r_{B}\neq0$), where the interplay between QOC and feedback takes place. Compared with Figs.\,\ref{Fig3}(a) and \ref{Fig3}(b), one can intuitively find that, under the same QOC strength, a nonzero value of $r_{B}$ (e.g., $r_{B}=0.2$) leads to a higher maximum value of $E_{\mathcal{N}}$, implying that COM entanglement can be further improved by QOC in the presence of coherent feedback. To further explore how the COM entanglement can be optimized by the synergistic control of QOC and coherent feedback, we plot the dependence of $E_{\mathcal{N}}$ on the corresponding controlling parameters $g_{2}/g_{1}$, $\bar\Delta/\omega_{m}$, $r_{B}$, and $\theta$ in Figs.\,\ref{Fig3}(c)-\ref{Fig3}(f). From these results, it is found that the synergetic control of COM entanglement is characterized by two generic features: (i) For a fixed value of $\theta$, $E_{\mathcal{N}}$ increases monotonically with both $r_{B}$ and $g_{2}/g_{1}$ and reaches its maximum value just at the instability threshold; (ii) $E_{\mathcal{N}}$ shows a periodic dependence on the optical phase shift $\theta$ with a period of $2\pi$, within which both maxima and minima occur. However, we emphasize that an excessively large $r_B$ may also drive the system into the unstable parameter region due to the modified stability boundary induced by the reduction of $\tilde{\kappa}$. These results are consistent with the system stability analysis shown in Figs.\,\ref{Fig2}(c)-\ref{Fig2}(f). In addition, compared with the case without QOC and feedback, the two extreme values are found to be either enhanced or suppressed. Here the achievable maximum value of $E_{\mathcal{N}}$ can be up to about $0.48$ under the optimized condition [see the yellow star in Fig.\,\ref{Fig3}(f)], which is about $12$ times larger than that obtained without QOC and coherent feedback. These results indicate that the synergistic control of QOC and feedback provides a versatile framework for manipulating COM entanglement, allowing both enhancement and switching of the COM entanglement.
\begin{figure}[t]
\centering
\includegraphics[width=0.48\textwidth]{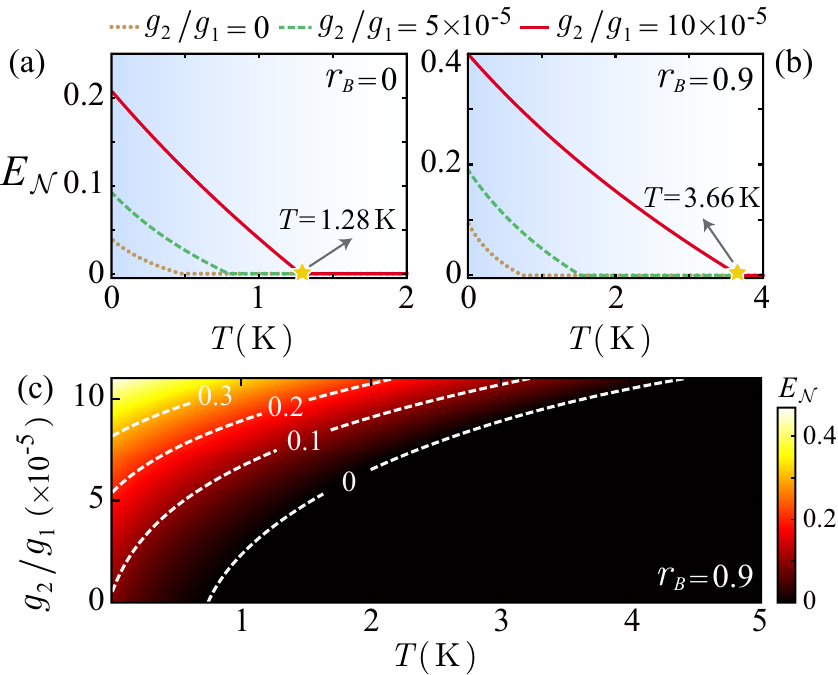}
\caption{Thermal robustness of optomechanical entanglement under the synergistic control of QOC and feedback. (a,b) Logarithmic negativity $E_{\mathcal{N}}$ as a function of the bath temperature $T$ for different values of QOC strength $g_{2}$, with $r_{B}=0$ in (a) and $r_{B}=0.9$ in (b). (c) Density plot of $E_{\mathcal{N}}$ as a function of the bath temperature $T$ and the QOC strength $g_{2}/g_{1}$. Here we take $\bar{\Delta}/\omega_{m}=0.25$, $\theta=3\pi/2$, and the other parameters are the same as in Fig.\,\ref{Fig3}.}\label{Fig4}
\end{figure}

Physically, the observed tunability and enhancement of COM entanglement stem from the effective control over the system stability, achieved through the synergistic effect of QOC and coherent feedback. This mechanism can be understood as follows. On one hand, following the analysis in Ref.~\cite{Genes2008PRA}, we note that the amount of entanglement $E_{\mathcal{N}}$ exhibits a monotonic dependence on the effective COM coupling strength $\tilde{G}$. Yet, for certain system parameters, the applicable maximum value of $\tilde{G}$ is physically limited by the system stability condition derived in Eq.~(\ref{eq:stability}). This constraint dictates that the steady-state COM entanglement is always maximized at the system's instability threshold. On the other hand, the threshold of the system stability itself is not immutable. In our model, the system stability under the red-detuned regime is governed by the condition $C_{1}$, which depends on the effective system parameters $\Omega_{m}$, $\tilde{\kappa}$, $\tilde{G}$, and $\tilde{\Delta}$. In the presence of QOC and coherent feedback, these parameters are rendered controllable via their interplay. As a result, by optimizing the controlling parameters, one can effectively shift the instability threshold, permitting the system to sustain a stronger COM coupling strength $\tilde{G}$ before entering unstable region. This allows for a higher maximum value of $E_{\mathcal{N}}$, even though accompanied by a widening of the instability region compared to the bare COM system.

Figure\,\ref{Fig4} further shows the thermal robustness of the optomechanical entanglement under the synergistic control of QOC and feedback. In Fig.\,\ref{Fig4}(a), we plot the logarithmic negativity $E_{\mathcal{N}}$ as a function of bath temperature $T$ in the absence of coherent feedback ($r_B = 0$). It is seen that, for the COM system only with LOC ($g_2=0$), the optomechanical entanglement can persist up to $T \approx 0.5\,\mathrm{K}$. When introducing QOC to the COM system, the entanglement robustness is improved, with e.g., its survival temperature extended to $T \approx 1.28\,\mathrm{K}$ for $g_2/g_1=10\times10^{-5}$ [see yellow star in Fig.\,\ref{Fig4}(a)]. Furthermore, a more significant enhancement of the thermal robustness can be observed when the coherent feedback is introduced. As shown in Fig.\,\ref{Fig4}(b), for $r_B=0.9$ and $\theta=0$, the entanglement survival temperature is remarkably enhanced compared to the case of $r_B=0$. Specifically, it is seen that, for $g_2/g_1=10\times 10^{-5}$, the optomechanical entanglement persists up to $T = 3.66\,\mathrm{K}$, which is nearly $3$ times higher than the case without feedback. To provide a more comprehensive view of this joint effect, we further present $E_{\mathcal{N}}$ as a function of both $T$ and $g_2/g_1$ in Fig.\,\ref{Fig4}(c) with $r_B=0.9$ and $\theta=0$, where the entanglement region (with $E_{\mathcal{N}}>0$) expands significantly toward the high-temperature regime as the QOC strength increases. By comparing the results in Figs.\,\ref{Fig4}(a)-\,\ref{Fig4}(c), it is found that the synergistic effect of QOC and coherent feedback allows the optomechanical entanglement to be robustly maintained for bath temperatures even at several Kelvins, which relaxes the thermal requirements for the generation and manipulation of such macroscopic entanglement and enhances its feasibility in current experimental platforms.

\begin{figure*}[htbp]
\centering
\includegraphics[width=0.96\textwidth]{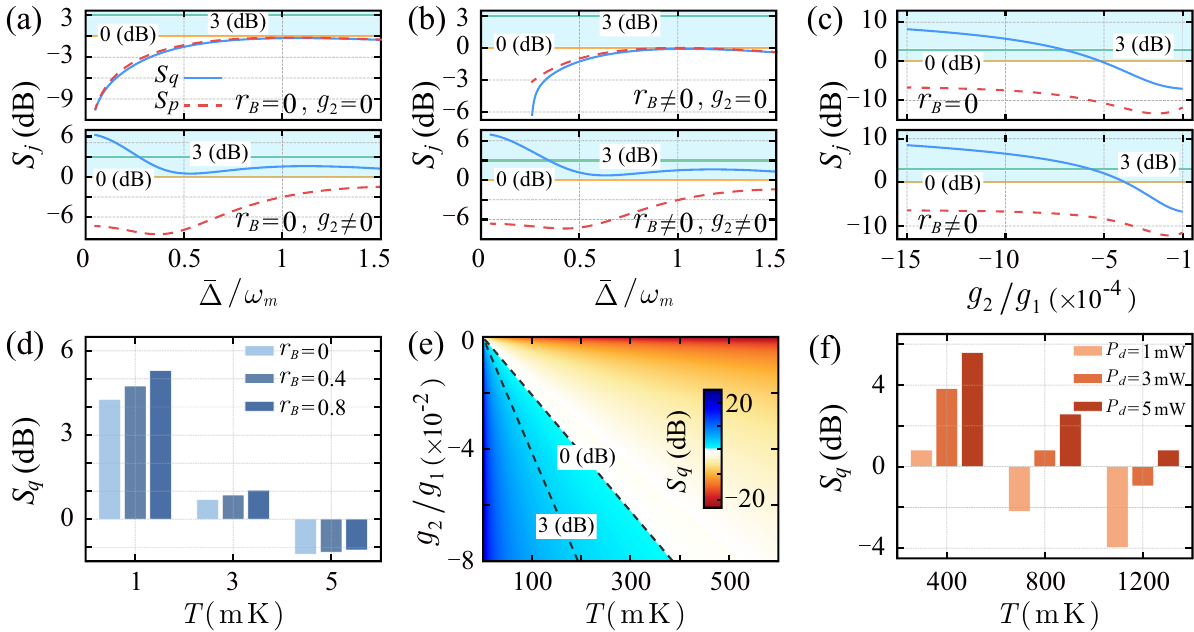}
\caption{Enhancement of mechanical quadrature squeezing via the synergistic effect of QOC and coherent feedback. The mechanical quadrature squeezing degree $S_{j}$ versus the scaled optical detuning $\bar{\Delta}/\omega_{m}$: (a) in the absence of coherent feedback ($r_{B}=0$) with $g_{2}/g_{1}=0$ (top panel) and $g_{2}/g_{1}=-10\times10^{-4}$ (bottom panel); (b) in the presence of coherent feedback ($r_{B}=0.8$, $\theta=0$) with $g_{2}/g_{1}=0$ (top panel) and $g_{2}/g_{1}=-10\times10^{-4}$ (bottom panel). (c) The mechanical quadrature squeezing degree $S_{j}$ as a function of dimensionless QOC strength $g_{2}/g_{1}$ for $\bar{\Delta}/\omega_{m}=0.1$ with $r_{B}=0$ (top panel) and $r_{B}=0.8$, $\theta=0$ (bottom panel). The blue solid and red dashed lines correspond to $S_{q}$ and $S_{p}$, respectively. (d) The squeezing degree $S_{q}$ as a function of bath temperature $T$ for different values of $r_{B}$, with $\theta=0$ and $g_{2}/g_{1}=-10\times10^{-4}$. (e) Density plot of $S_{q}$ as a function of bath temperature $T$ and QOC strength $g_{2}/g_{1}$, with $r_{B}=0.8$, $\theta=0$, and $\bar{\Delta}/\omega_{m}=0.2$. (f) The squeezing degree $S_{q}$ as a function of bath temperature $T$ for different values of $P_{d}$, with $r_{B}=0$ and $g_{2}/g_{1}=-8\times10^{-2}$. The parameters used here are the same as in Fig.\,\ref{Fig3} except for $\kappa_1=2\pi\times2.25\,\mathrm{MHz}$, $\kappa_2=2\pi\times0.75\,\mathrm{MHz}$, $P_{d}=1.6\,\mathrm{mW}$, and $T=1\,\mathrm{mK}$.}\label{Fig5}
\end{figure*}

\section{The synergistic effect of QOC and coherent feedback on steady-state mechanical squeezing}\label{secIV}
The synergistic control of QOC and coherent feedback also provides a viable route to achieving strong mechanical quadrature squeezing beyond the $3$dB limit. To quantify the squeezing of the membrane motion, we define the degree of squeezing as (in units of dB)~\cite{Zhang2019PRA}
\begin{align}
S_j=-10\log_{10}\left(\frac{\sigma_j}{\sigma_{\rm zpf}}\right),
\end{align}
where $\sigma_j$ ($j=q,p$) is the variance of the mechanical quadrature operator, obtained from the corresponding diagonal elements of the CM $V$, and $\sigma_{\rm zpf}=|\langle[\delta\hat{q},\delta\hat{p}]\rangle|/2=1/2$ denotes the zero-point fluctuation of the membrane motion. $S_j>0$ indicates that the fluctuation of the $j$ quadrature is squeezed. In Particular, $S_j>3$ corresponds to a $50\%$ reduction of noise below the zero-point fluctuation, i.e., $\sigma_j<\sigma_{\rm zpf}/2$, which is regarded as strong mechanical squeezing beyond the $3$dB limit.

In Figs.\,\ref{Fig5}(a) and \ref{Fig5}(b), we present the squeezing degrees $S_{q}$ and $S_{p}$ as functions of the scaled optical detuning $\bar\Delta/\omega_{m}$ for different reflection coefficients $r_{B}$ and QOC strengths $g_{2}/g_{1}$. The results highlight the distinct roles of QOC and coherent feedback. Specifically, in the absence of QOC ($g_{2}/g_{1}=0$), no mechanical squeezing is present ($S_{q}<0$ and $S_{p}<0$), regardless of whether coherent feedback is applied. In contrast, for $g_{2}/g_{1}\neq 0$, the fluctuation of the $q$ quadrature exhibits squeezing ($S_{q}>0$), and even without feedback ($r_{B}=0$), the $3$dB limit of squeezing can be beaten, with $S_{q}>3$ over a finite range of detuning $\bar\Delta$. The presence of feedback ($r_{B}\neq 0$) can slightly enhance the squeezing degree $S_q$ while maintaining the overall profile of the squeezing spectrum. To further present the crucial role of QOC, we show the behavior of $S_{q}$ versus QOC strength $g_{2}/g_{1}$ for different reflection coefficients $r_{B}$ in the vicinity of $\bar\Delta/\omega_{m}=0.1$ in Fig.\,\ref{Fig5}(c). As is seen, for the negative sign of $g_{2}/g_{1}$, $S_{q}$ can exceed the $3$dB limit in a broad parameter region with optimal value above $9$dB, and the incorporation of feedback ($r_{B}\neq 0$) allows the system to achieve the same degree of squeezing with a smaller required magnitude of $g_2$. Notably, it is also confirmed that for the positive sign of $g_{2}/g_{1}$, mechanical squeezing is unachievable in both quadratures with $S_{q}<0$ and $S_{p}<0$. These results suggest that for mechanical quadrature squeezing generation, QOC serves as an essential fundamental resource, whereas feedback acts as a supplementary tool to optimize the squeezing performance and lower the requirement for QOC strength.

Physically, the enhancement of mechanical squeezing originates from the synergistic regulation of the mechanical susceptibility and system stability induced by the interplay between QOC and coherent feedback. This mechanism can be understood as follows. On one hand, mechanical squeezing is closely associated with the amplification of the mechanical response to quantum fluctuations, which is characterized by the mechanical susceptibility $\chi_m(\omega)=[\Omega_m^2-\omega^2-i\omega\gamma_m]^{-1}$. In the presence of negative QOC ($g_2<0$), the effective mechanical frequency $\Omega_m=\omega_m+2g_2|\alpha_s|^2$ decreases, leading to a softened mechanical mode with $\Omega_m<\omega_m$. Such a reduction in $\Omega_m$ significantly enhances the low-frequency mechanical susceptibility, thereby making the mechanical resonator more sensitive to the quantum fluctuations responsible for generating squeezing. Consequently, the achievable degree of mechanical squeezing can be considerably enhanced for negative QOC strength. On the other hand, the enhancement of squeezing through mechanical softening is intrinsically constrained by the system stability condition, because the reduced $\Omega_m$ tends to drivie the system toward the unstable regime. In our model, however, coherent feedback provides a flexible tool to compensate the reduction in $\Omega_m$ via the tunable parameters $\tilde{\kappa}$ and $\tilde{\Delta}$ [cf. $C_1$ in Eq.~(\ref{eq:stability})]. As a result, the system can safely operate in a parameter region with simultaneously enhanced susceptibility and strong effective coupling, which thereby enables the presence of strong mechanical squeezing beyond the $3$dB limit.

Figs.\,\ref{Fig5}(d)-\ref{Fig5}(f) further shows the thermal robustness of the mechanical squeezing under different control parameters. In Fig.\,\ref{Fig5}(d), we compare the squeezing degree $S_q$ at different bath temperatures $T$ for $g_2/g_1 = -10 \times 10^{-4}$. It is seen that for weak QOC strength, mechanical squeezing typically requires relatively low temperatures (in the $\mathrm{mK}$ range) to be sustained. At a fixed temperature, the incorporation of feedback ($r_B > 0$) consistently enhances $S_q$, indicating that feedback remains a beneficial tool for optimizing squeezing performance under thermal noise. Moreover, as shown in Fig.\,\ref{Fig5}(e), a more significant improvement in thermal robustness can be achieved by increasing the QOC strength. It is seen that the survival temperature for mechanical squeezing ($S_q > 0$) significantly expands as $g_2$ increases. For instance, in terms of $g_2/g_1 = -8 \times 10^{-2}$, mechanical squeezing can persist up to $T \approx 380$ mK, and the $3$ dB limit ($S_q > 3$) can be surpassed for temperatures below $200$ mK. In addition, as shown in Fig.\,\ref{Fig5}(f), the driving power $P_d$ also plays a crucial role in improving the thermal robustness of mechanical squeezing. It is seen that, for $g_2/g_1 = -8 \times 10^{-2}$, the survival temperature of mechanical squeezing is enhanced when $P_d$ increases. Notably, it is seen that mechanical squeezing still remains achievable even at $T = 1.2\,\mathrm{K}$ for $P_d = 5\,\mathrm{mW}$. By comparing the results in Figs.\,\ref{Fig5}(d)-\ref{Fig5}(f), it can be intuitively found that while increasing $r_B$, $g_2$, and $P_d$ all contribute to the improvement of the thermal robustness of mechanical squeezing, the QOC strength $g_2$ and the driving power $P_d$ plays a more important role. This is because the degree of mechanical squeezing is related to the effective COM coupling strength $\tilde{G}$, which can be directly enhanced by both $g_2$ and $P_d$. These synergistic effects effectively raise the survival bath temperature for mechanical squeezing, which ensures the experimental feasibility of generating and utilizing such robust mechanical squeezing in practice.

\section{Analysis of Experimental Feasibility}\label{secV}
In our calculations, we choose the following default parameters similar to those used in~\cite{Sheng2020PRL}: cavity resonant frequency $\omega_c/2\pi=370\,\textrm{THz}$, mechanical frequency $\omega_m/2\pi=10\,\textrm{MHz}$, cavity decay rates $\kappa_1=\kappa_2=2\pi\times1.5\,\textrm{MHz}$, mechanical damping rate $\gamma_m/2\pi=100\,\textrm{Hz}$, and driving wavelength $\lambda_{d}=810\,\textrm{nm}$ with power $P_{d}=0.8\,\textrm{mW}$. Note that, in order to work in the resolved sideband limit ($\kappa<\omega_m$), we here increased the mechanical frequency by approximately an order of magnitude to $10\,\textrm{MHz}$, while choosing a relatively lower mechanical quality factor $Q_m=\omega_m/\gamma_m=10^5$ compared to Ref.~\cite{Sheng2020PRL}, which ensures that $Q_m$ remains within a practical experimental range for dielectric membrane resonators~\cite{Sheng2020PRL,Thompson2008Nature,Sankey2010NP}. The single-photon LOC strength $g_{1}$ is estimated by $g_{1}=(\partial\omega_{c}/\partial x)x_{\textrm{zpf}} \approx -1351.38\,\textrm{Hz}$, with $x_{\textrm{zpf}}=\sqrt{\hbar/m\omega_m}$ the zero-point fluctuation of the mechanical displacement. Given that QOC is typically orders of magnitude weaker than LOC~\cite{Thompson2008Nature,Sankey2010NP,Vanner2011PRX,Flowers2012APL}, we thus vary the ratio $g_{2}/g_{1}$ from $10^{-5}$ to $10^{-2}$ in our simulations, and the sign of $g_{2}$ can be flexibly determined by the membrane's equilibrium position $q_{0}$ [see the expression of $g_{2}$ in Eq.(\ref{eq:g12})]. In addition, the thermal robustness of our proposed scheme is verified in Figs.~\ref{Fig4} and \ref{Fig5}, which show that the synergistic effect of QOC and coherent feedback improves the robustness of both optomechanical entanglement and mechanical squeezing. Particularly, it is found that considerable entanglement and squeezing can be generated at sub-Kelvin temperatures (hundreds of $\textrm{mK}$), with their survival temperatures even persisting up to a few Kelvin. Such bath temperatures are experimentally feasible via cryogenic precooling in dilution refrigerators. For example, recent experiments demonstrated that dielectric membrane resonators can be precooled below $T=300\,\textrm{mK}$ by dilution refrigerators~\cite{Thompson2008Nature,Zwickl2008APL}. Furthermore, note that the thermal phonon occupation $\bar{n}_{m}$ is inversely proportional to the mechanical frequency, our proposed scheme could be further extended to higher-frequency mechanical systems to achieve better thermal robust. For example, if QOC can be realized in systems incorporating molecular or exciton mechanical oscillators~\cite{Zou2024PRL,Huang2024PRB,Huang2025PRA,Yu2026PRL}, the synergistic control could potentially be applied to engineer quantum effects even at room temperatures.

\section{Conclusion}\label{secVI}

In summary, we have investigated the synergistic effects of QOC and coherent feedback on the generation and manipulation of COM entanglement and mechanical squeezing. Specifically, we consider here a membrane-embedded COM system coupled with a coherent feedback loop, in which the membrane interacts with the cavity mode through both LOC and QOC. This hybrid architecture offers two critical advantages over a bare COM system in terms of parameter tunability: (i) the incorporation of QOC provides a flexible means to tune the frequency of the membrane, where the sign of QOC determines whether the mechanical mode becomes stiffer or softer; (ii) feeding the output field back into the cavity effectively reduces the cavity decay rate. Thanks to these advantageous features, we showcase that for suitable QOC and feedback parameters, not only the degree of both COM entanglement and mechanical squeezing can be considerably enhanced, their thermal robustness are simultaneously improved as well. This enhancement of quantum coherence originates from the strategic reconfiguration of the system's stability condition enabled by the synergistic interplay of QOC and feedback, which relies on the all-optical approaches without the use of sophisticated quantum engineering techniques such as backaction-evading measurements~\cite{Szorkovszky2011PRL} or reservoir engineering~\cite{Rabl2004PRB}. Moreover, although we have considered here a specific case of a dielectric-membrane-embedded COM system, we note that the proposed synergistic scheme remains applicable to a broad range of experimental platforms with both LOC and QOC, including those involving trapped cold atoms~\cite{Purdy2010PRL} or microspheres~\cite{Li2011NP}, double-disk structures~\cite{Lin2009PRL}, and magnomechanical devices~\cite{Makinen2025NC}. In a broader view, COM systems featuring both LOC and QOC have been found to host a variety of unique physical properties, such as enabling continuous time crystals coupled to mechanical modes~\cite{Makinen2025NC}. As such, we envision that future developments incorporating our proposal with such unique properties could enable more versatile degrees of freedom for manipulating light-motion interaction, which could further offer a promising tool for engineering diverse nonclassical states~\cite{Rabl2011PRL,Lu2018PRAPP,Zhu2019PRA,Simon2018PRL,Sheng2020PRL} based on COM devices.

\section{Acknowledgment}
Y.-F.J. is supported by the the National Natural Science Foundation of China (NSFC, Grant No.~12405029) and the Natural Science Foundation of Henan Province (Grant No.~252300421221). H.J. is supported by the NSFC (Grants No.~11935006, 12421005), the National Key R$\&$D Program of China (Grant No.~2024YFE0102400), and the Hunan Provincial Major Sci-Tech Program (Grant No.~2023ZJ1010). L.-M.K. is supported by the NSFC (Grants No.~12247105, 11935006, 12175060 and 12421005), the Hunan Provincial Major Sci-Tech Program (Grant No.~2023ZJ1010), and the Henan Science and Technology Major Project (Grant No.~241100210400). H.Z. is supported by the Natural Science Foundation of Henan Province (Grants No.~252300421794, 242300420665) and the Doctoral Research Foundation of Zhengzhou University of Light Industry (Grant No.~2022BSJJZK20). Y.-C.L. is supported by the NSFC (Grant No. 12304474) and the Natural Science Foundation of Henan Province (Grant No. 252300421222). Y.W. is supported by the NSFC (Grant No.~12205256).


%

\end{document}